# Simple variation of the logistic map as a model to invoke questions on cellular protein trafficking[*]

Sepehr Ehsani

Department of Laboratory Medicine and Pathobiology and Tanz Centre for
Research in Neurodegenerative Diseases, University of Toronto

Toronto, Ontario, Canada

sepehr.ehsani@utoronto.ca

24 June 2012

Many open problems in biology, as in the physical sciences, display nonlinear and 'chaotic' dynamics, which, to the extent possible, cannot be reasonably understood. Moreover, mathematical models which aim to predict/estimate unknown aspects of a biological system cannot provide more information about the set of biologically meaningful (e.g., 'hidden') states of the system than could be understood by the designer of the model *ab initio*. Here, the case is made for the utilization of such models to shift from a 'predictive' to a 'questioning' nature, and a simple natural-logarithm variation of the logistic polynomial map is presented that can invoke questions about protein trafficking in eukaryotic cells.

## INTRODUCTION

Our (human) perception of the natural world around us (i.e., the 'real' world) is mediated through five sensory systems which provide a continuous stream of information for a working, generative, model of the world in the brain (see for example [1, 2]). It therefore follows that our perception and, consequently, cognition, must be somehow bounded by the senses [3]. These cognitive limits are not well-defined, but have begun to be understood in areas such as the inherent faculty of language [4]. Before attempts at understanding these limits start to bear fruit, modeling the outside world cannot come close to reality (**Fig. 1**, and see [5] for a perspective). This situation can be compared to a game of association football, in which the players must have a good sense of the boundaries of the field before any meaningful attempt at playing the game can be made. It is therefore no surprise that many detailed observations of the 'real' world, from a simple pendulum to the population dynamics of a colony of ants, seem 'complex' and comprise 'random' features [6]. If the observed effects were truly (inherently) random, the natural system would not function, given that, by definition, a random nature could produce infinite functionalities, which is equal to no functionality at all in one system. Moreover, if an existential entity were truly random, it would be 'nothing,' but given that it is 'something' (i.e., it exists), it may only be random to the observer. Furthermore, and surprisingly, increasing the complexity of the starting conditions of an experimental paradigm can, at times, lead to much simpler outcomes [7].

---

[*] The first draft of this paper was prepared as part of the Santa Fe Institute's Complex Systems Summer School, June 2012.

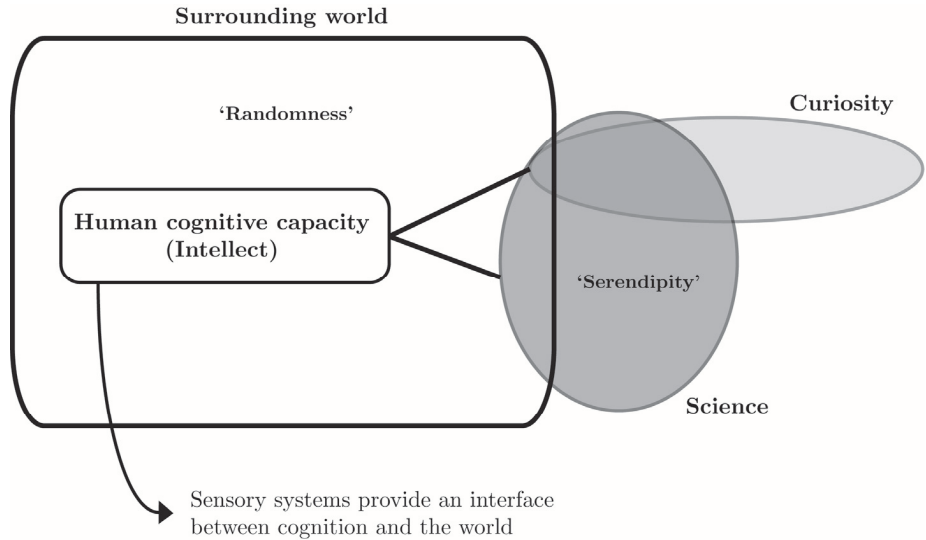

**Figure 1. Limits of understanding and the outside world.** In this schematic representation, the sensory systems are imagined to provide an interface between the human cognitive capacity and aspects of the outside world, and to enable a metaphorical 'decoupling' transition between the 'randomness' of the outside (hidden states) and the perceived order of the inside (observable, albeit 'complex', states). Cognition can then wander into scientific questions (i.e., rational inquiry), which may arise (i) out of curiosity or (ii) serendipitously. At times, these inquiries lead to a better understanding of (or hypothesis about) an aspect of the surrounding world.

Modeling natural phenomena is helpful in presenting the known facets of a system in a computationally formulable framework, but cannot be expected to provide serious predictions about or estimations of meaningful ('hidden') states of the system (see for example [8]). This, nevertheless, does not (and should not) put a stop on the development of more complex predictive algorithms; however, such methodologies have the potential to become increasingly incoherent with the actual physical phenomenon they try to predict, and may appear more like 'mathematical poetry', pursued for their own sake rather than their use as real tools for understanding. An alternative usage of such algorithmic methodologies is to modify them into question-generating mechanisms, whereby their output can invoke previously neglected questions. To that end, a simple variation of a polynomial map is presented here as a means for asking questions on an aspect of cellular biology, wherein cells act as manipulable natural units of complexity.

## MODELS AND RESULTS

We begin our model by tabulating the dimensions of a model prokaryotic (*Escherichia coli*) and eukaryotic (*Schizosaccharomyces pombe*) cell (**Table 1**). For comparison, the diameter of a membrane-bound protein, a dimer of the cellular prion protein ($PrP^C$), is also provided. Given the lack of (i) a nuclear membrane, (ii) endoplasmic reticulum compartments, and (iii) spatial (and possibly temporal) separation between transcription and translation in prokaryotes [9], *S. pombe*'s dimensions will be used for the models presented herein (**Fig. 2**).

**Table 1. Dimensional features in a model prokaryote, eukaryote and membrane protein.**

|  | **Prokaryote (*E. coli*)** | **Eukaryote (*S. pombe*)** | **Prion protein (PrP$^C$)** |
|---|---|---|---|
| **Volume** | 0.6-0.7 μm$^3$ [10] | 54.3 μm$^3$ [11] | N/A |
| **Diameter** | 0.5-1.0 μm [9] | 10.0 μm$^\dagger$ [12] | ~100 Å$^\#$ [13] |
| **Nuclear diameter** | N/A | 3.0 μm$^\ddagger$ [12] | N/A |

$^\dagger$This value corresponds to the length of the cell, as *S. pombe* is elliptical.
$^\ddagger$The nuclear diameter is assumed to be approximately the same as the width of the cell in *S. pombe*.
$^\#$This is an approximate maximum length of a dimer of the prion protein.

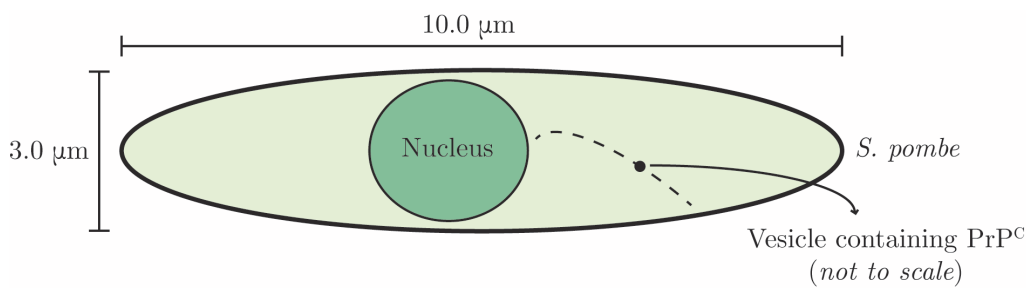

**Figure 2. Schematics of *S. pombe* based on values presented in Table 1.** For simplicity, the path of a protein-containing vesicle is assumed to start from the vicinity of the nucleus to the plasma membrane for a distance of approximately a third of the diameter of the elliptical cell.

Proteins are the main functional agents in cells, and they have to be localized to their specified cellular compartment(s) in a timely mode after translation on the endoplasmic reticulum. This process, called protein trafficking, is mediated by protein-containing vesicles which utilize the cytoplasmic microtubular network to reach, in the case of membrane proteins that will be the focus of this study, the cell periphery and fuse with the 20-nm-wide plasma membrane [14, 15]. A prion protein dimer, for example, travels a length approximately 300-times its length (100 Å vs. a third of *S. pombe*'s diameter of 10 μm) in a vesicle [14] to the membrane. This journey, however, is through a very crowded cytosolic environment [16], with a multitude of unknowns.

The most rudimentary mode to explain this transport is through diffusion. In this view, high concentrations of a protein produced on the endoplasmic reticulum diffuse to locales of lower concentration on the periphery of the cell. An example of such a model is demonstrated (**Fig. 3**) using NetLogo's *Diffusion on a Directed Network* model [17, 18]. This model, however, suffers from the significant fact that cytosolic and membranal locations in a eukaryotic cell are compartmentalized [19], and therefore the system is not 'continuous'; moreover, although some may consider the nuclear and cytosolic compartments to be more connected, recent evidence suggests that the discontinuity may be more than previously thought [20]. This diffusion methodology, therefore, is not continued further.

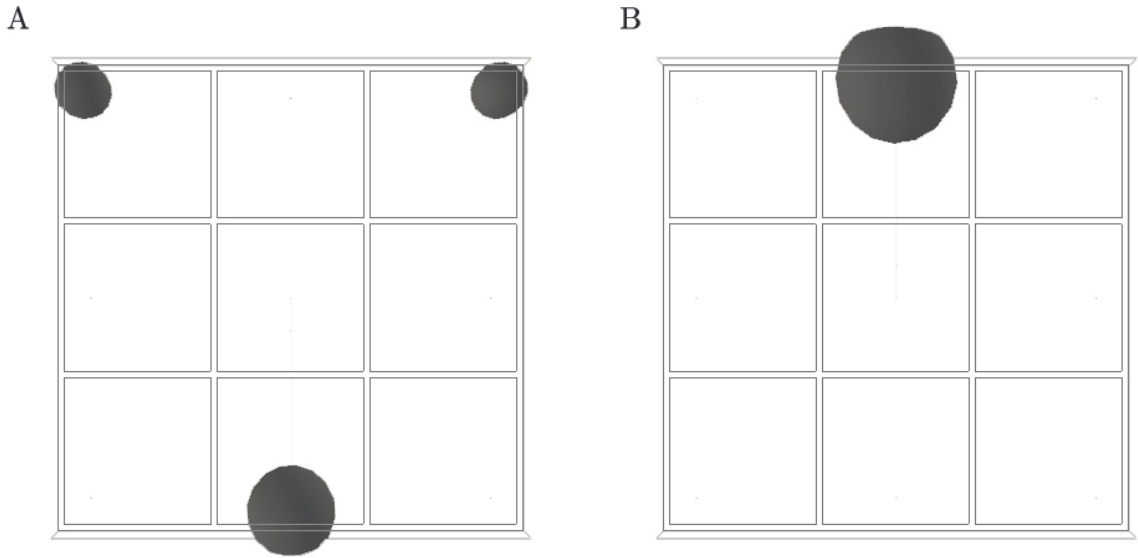

**Figure 3. Diffusion model.** This *Diffusion on a Directed Network* model assumes high concentrations of a new translated protein close to the nucleus and lower concentrations in the periphery (membrane) in a grid-like network (**A**). Over time, the protein diffuses to a single location on the periphery where the protein is needed on the membrane, possibly even different from the location of the low-concentration populations already on the periphery (**B**).

We can then consider the logistic map [21], a ubiquitous polynomial map with chaotic potential:

$$x_{n+1} = rx_n(1 - x_n) \quad \text{Equation 1}$$

If $x_n$ is the current position of a protein-containing vesicle, and $x_{n+1}$ is its next position in a discrete time frame, if we set $r = 3$, a fast-rising transport followed by constrained oscillation is observed (**Fig. 4A**). In this model, the fast rise can be assumed to represent the vesicle's movement toward the membrane (**Fig. 4B**), and possess a mean movement rate of ~0.5 µm/s through the cytosol [22]. The following oscillations can represent the movement of the protein on the membrane, such as when embedded on lipid rafts on the phospholipid bilayer, although the velocity of protein-containing lipid rafts is not unambiguously determined [23].

If a protein does not localize to lipid rafts and remains approximately stationary on the membrane (similar to models described previously for other cellular processes [24]), the logistic map can be modified slightly to represent the stationary phase following a rapid rise (**Fig. 5**):

$$x_{n+1} = \sin(rx_n(1 - x_n)) \quad \text{Equation 2}$$

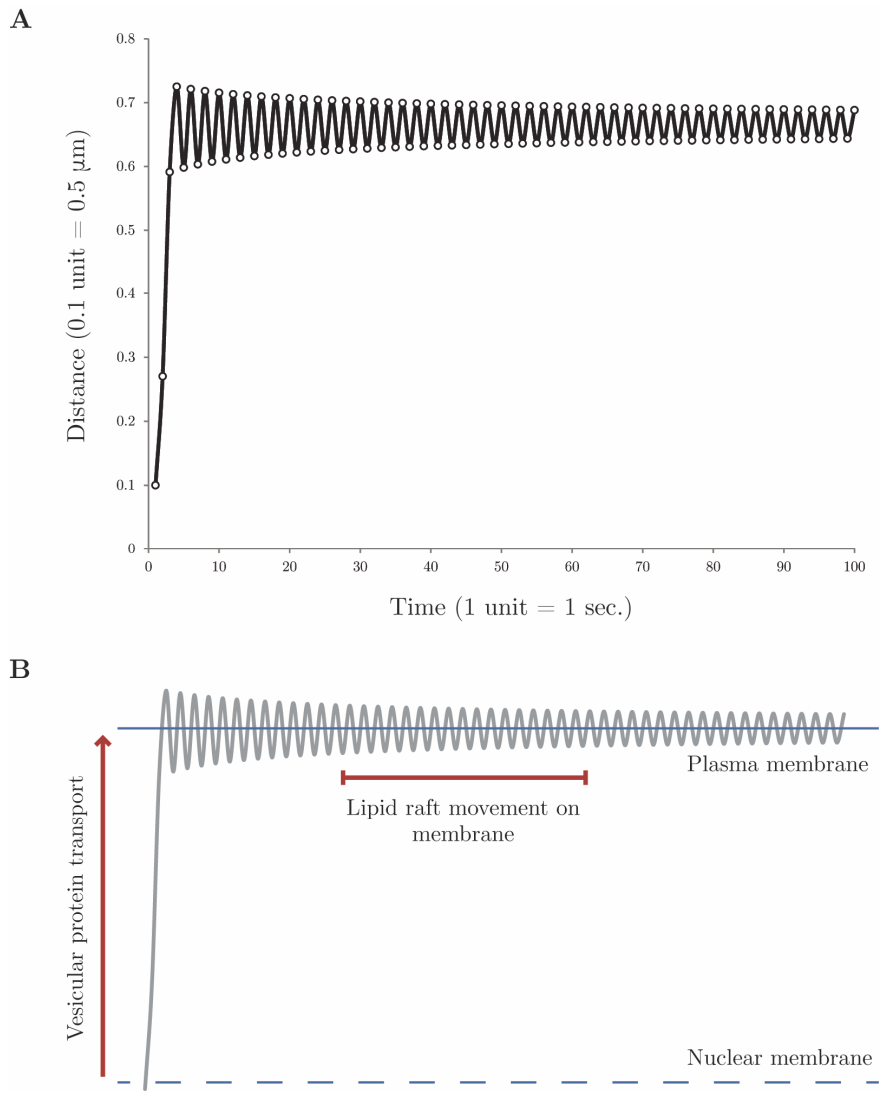

**Figure 4. Unmodified logistic map model ($r = 3$) of protein trafficking to the plasma membrane.** (**A**) Based on a vesicle movement rate of ~0.5 μm/s, the rise of the curve can represent a protein-containing vesicle's transport to the membrane, and the ensuing oscillation can represent the movement of the protein on the membrane on a lipid raft. These assumptions are annotated in (**B**).

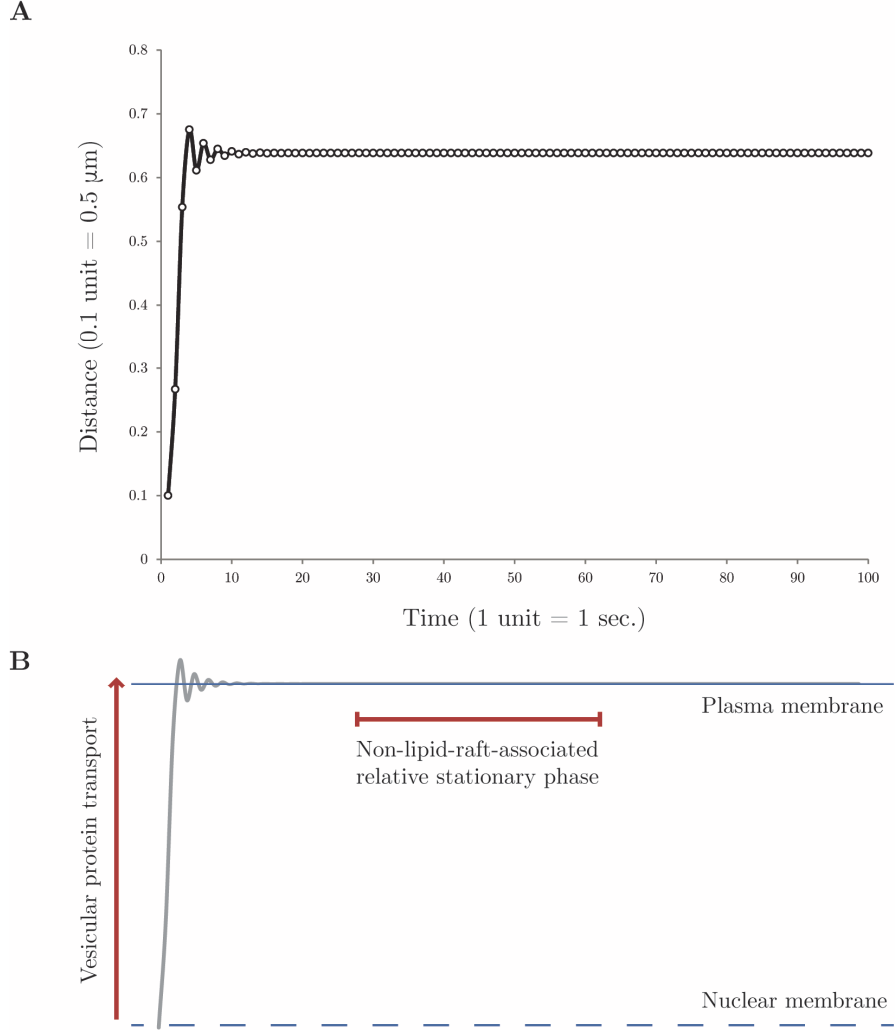

**Figure 5. Sine-function-modified logistic map ($r = 3$) for membrane-stationary proteins.**
(**A**) With a sine-function modification, the protein now adjusts to a stationary phase after the rise to the membrane (rate of ~0.5 μm/s) in the logistic model. Annotations appear in (**B**).

To represent a continuous cycle of (i) protein trafficking to the membrane, (ii) endocytosis of the membrane protein and (iii) protein degradation, followed by (iv) another cycle of protein translation, a natural-logarithm of the absolute-value of the logistic map is presented, where again $r = 3$ (**Fig. 6**):

$$x_{n+1} = \ln|rx_n(1 - x_n)| \qquad \textbf{Equation 3}$$

This bears some resemblance to the Lyapunov exponent of chaotic systems, which has been discussed previously as calculable by "averag[ing] … the natural logarithm of the absolute value of the derivatives of the map function evaluated at the trajectory points" [25].

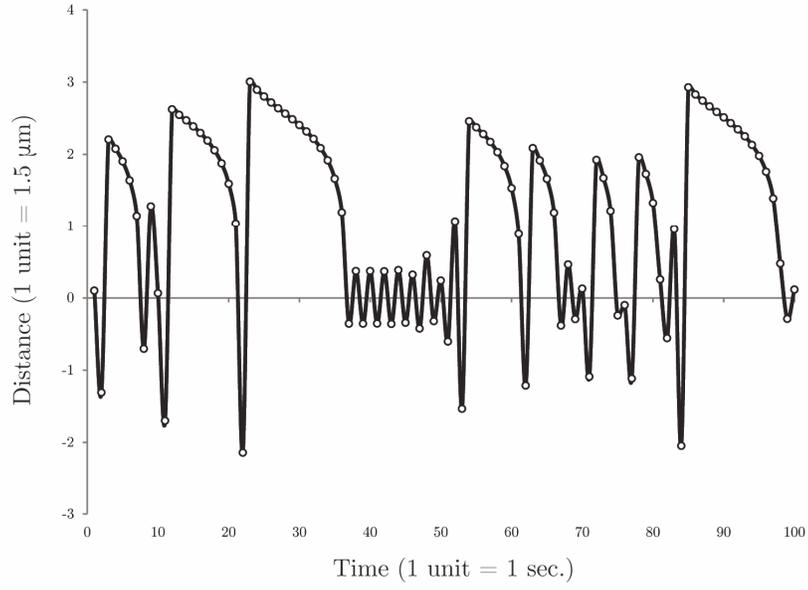

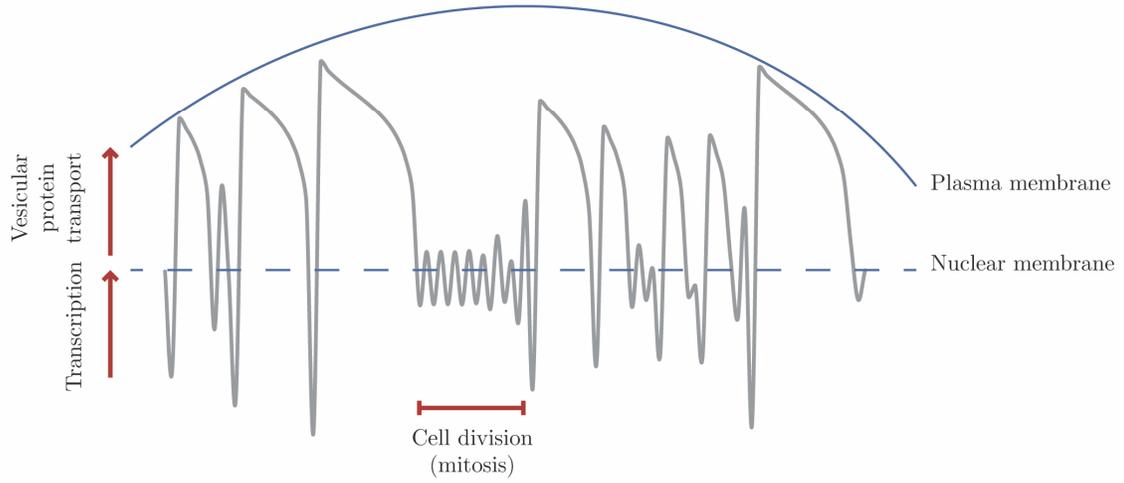

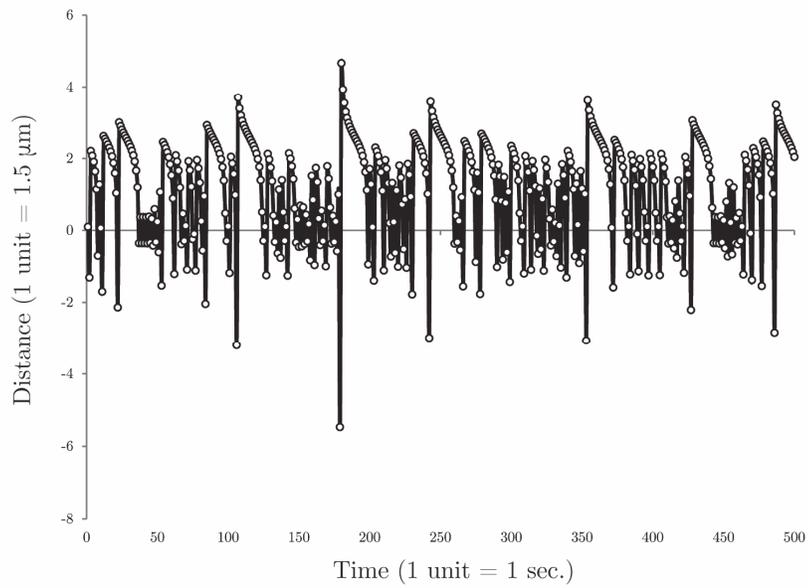

**Figure 6. Natural-logarithm variation of the logistic map ($r = 3$) representing a continuous model of protein trafficking and degradation.** (**A**) In this extended 'chaotic' model, cycles of continuous protein trafficking to the membrane, followed by degradation, take place to varying degrees and are then superseded by a period of oscillation at the baseline, representative, for example, of cellular division (mitosis). Each protein degradation cycle culminates in a dip, which can be a metaphorical representation for renewed transcription in the nucleus. (**B**) An annotated version of (**A**). (**C**) The protein trafficking/degradation cycle continues a pseudo-random periodicity over a longer time range. Please note that the x-axis time unit of the graphs corresponds to the cytosolic trafficking of a protein-containing vesicle to the membrane only, and cannot account for the variable residency times of different proteins on the membrane [26], or for the life cycle of different cell types.

## DISCUSSION

It is a truism that our understanding of seemingly simple natural phenomena is far too little to allow for a more or less comprehensive incorporation of all existing state variables in a predictive model. As an example, one can point to the field of fluid mechanics, where, given that the different interactional models of seemingly simple water molecules are still very challenging problems [27, 28], how can one expect a somewhat realistic predictive algorithm of more complex aqueous fluids? It follows that the ability of today's very capable mathematical/computational modeling algorithms [29, 30] to provide novel insights into the nature of a biological cell must also be constrained, or at times even misleading, in view of the inevitable dependence of models on the current domain of biological knowns, which form an extremely limited set.

As manifested by the original purpose of modeling in the natural sciences [31], the simple variations of the logistic map presented here, for example, can lead to some elementary, yet practical, questions on the process of protein trafficking in eukaryotic cells. Some of these research questions could be:

(i) How many times is a given amino acid recycled in a cell [32]? The solution can also act as an indirect measure towards the quantification of protein degradation [33]. Moreover, theoretically, can an amino acid, passed through successive cycles of cell division, be recycled indefinitely? In other terms, how efficient is cellular biology in recycling amino acids? The pseudo-random patterns generated by the logarithmic variation of the logistic map (**Fig. 6**) suggest a notion of contextually-bound determinism.

(ii) Using the current methods of single-molecule tracking on the plasma membrane [34, 35], can one measure the velocity of lipid rafts [36] while determining (a) whether the rafts move in certain domains on the membrane and (b) if the velocity converges to a relatively constant absolute value (**Fig. 4**)?

(iii)    Based on a few instances of incomplete trafficking-degradation cycles observed in the modified chaotic map (**Fig. 6**), can it be determined whether all vesicular transports of a given protein pool to the membrane are successful? Is the failure rate relatively low, and if so, how is it achieved?

(iv)    To what extent do lipid rafts confer increased mobility to proteins located on them? Comparing models in **Figs. 4** and **5**, is the movement of non-lipid-raft proteins significantly lower than those on lipid rafts, or do the inherent dynamics of the phospholipid bilayer provide them with other movement modalities?

(v)    Do all vesicle-mediated transports of a given set of proteins happen with the same pace? Furthermore, is this pace influenced solely by the interactions of the vesicles with the microtubular structure, or could it be additionally modified by extracellular cues?

Instances exist in the biological literature in which such shifts in the utilization of models are already in practice. For example, a recent model of the emergence of carbon-fixation [37] has been presented in the context of new questions and biochemical experiments concerning metabolic pathways and their connections across different phyla. Likewise, phase transition models have been hypothesized to lead to testable questions regarding central nervous system arousal [38]. Finally, models to describe protein folding transition path times can lead to fundamental new questions about the phenomenon of protein folding in cells [39].

## REFERENCES

1.    Dyson, B.J. The advantage of ambiguity? Enhanced neural responses to multi-stable percepts correlate with the degree of perceived instability. *Front Hum Neurosci*, 2011. **5**: p. 73.

2.    Conant, R.C. and W.R. Ashby. Every good regulator of a system must be a model of that system. *Int J Systems Sci*, 1970. **1**(2): p. 89-97.

3.    Ehsani, S. Consciousness and subjective time: a plausible auditory approach. *Nature Precedings*, 2011.

4.    Chomsky, N. The Mysteries of Nature: How Deeply Hidden? *J Phil.*, 2009. **106**(4): p. 167-200.

5.    Buchanan, M. Know your limits. *Nat Phys*, 2012. **8**: p. 439.

6.    Crutchfield, J.P. Between order and chaos. *Nat Phys*, 2012. **8**: p. 17-24.

7.    Szostak, J.W. An optimal degree of physical and chemical heterogeneity for the origin of life? *Philos Trans R Soc Lond B Biol Sci*, 2011. **366**(1580): p. 2894-901.

8. Stumpf, M.P. and M.A. Porter. Mathematics. Critical truths about power laws. *Science*, 2012. **335**(6069): p. 665-6.

9. Nanninga, N. Morphogenesis of Escherichia coli. *Microbiol Mol Biol Rev*, 1998. **62**(1): p. 110-29.

10. Kubitschek, H.E. Cell volume increase in Escherichia coli after shifts to richer media. *J Bacteriol*, 1990. **172**(1): p. 94-101.

11. Zheng, T., et al. Quantitative 3D imaging of yeast by hard X-ray tomography. *Microsc Res Tech*, 2012. **75**(5): p. 662-6.

12. Neumann, F.R. and P. Nurse. Nuclear size control in fission yeast. *J Cell Biol*, 2007. **179**(4): p. 593-600.

13. Knaus, K.J., et al. Crystal structure of the human prion protein reveals a mechanism for oligomerization. *Nat Struct Biol*, 2001. **8**(9): p. 770-4.

14. Peters, P.J., et al. Trafficking of prion proteins through a caveolae-mediated endosomal pathway. *J Cell Biol*, 2003. **162**(4): p. 703-17.

15. Schmoranzer, J. and S.M. Simon. Role of microtubules in fusion of post-Golgi vesicles to the plasma membrane. *Mol Biol Cell*, 2003. **14**(4): p. 1558-69.

16. Harrison, P.M., et al. A question of size: the eukaryotic proteome and the problems in defining it. *Nucleic Acids Res*, 2002. **30**(5): p. 1083-90.

17. Wilensky, U. *NetLogo*. 1999; Available from: http://ccl.northwestern.edu/netlogo.

18. Stonedahl, F. and U. Wilensky. *NetLogo Diffusion on a Directed Network model*. 2008; Available from: http://ccl.northwestern.edu/netlogo/models/DiffusiononaDirectedNetwork.

19. Enns, C. Overview of protein trafficking in the secretory and endocytic pathways. *Curr Protoc Cell Biol*, 2001. **Chapter 15**: p. Unit 15 1.

20. Montpetit, B. and K. Weis. Cell biology. An alternative route for nuclear mRNP export by membrane budding. *Science*, 2012. **336**(6083): p. 809-10.

21. May, R.M. Simple mathematical models with very complicated dynamics. *Nature*, 1976. **261**(5560): p. 459-67.

22. Park, J.J., N.X. Cawley, and Y.P. Loh. Carboxypeptidase E cytoplasmic tail-driven vesicle transport is key for activity-dependent secretion of peptide hormones. *Mol Endocrinol*, 2008. **22**(4): p. 989-1005.

23. Garcia-Manyes, S., et al. Nanomechanics of lipid bilayers: heads or tails? *J Am Chem Soc*, 2010. **132**(37): p. 12874-86.

24. Hlavacek, W.S. and M.A. Savageau. Completely uncoupled and perfectly coupled gene expression in repressible systems. *J Mol Biol*, 1997. **266**(3): p. 538-58.

25. Yang, D., G. Li, and G. Cheng. Convergence analysis of first order reliability method using chaos theory. *Comput Struct*, 2006. **84**: p. 563-71.

26. Boisvert, F.M., et al. A quantitative spatial proteomics analysis of proteome turnover in human cells. *Mol Cell Proteomics*, 2012. **11**(3): p. M111 011429.

27. Perez, C., et al. Structures of cage, prism, and book isomers of water hexamer from broadband rotational spectroscopy. *Science*, 2012. **336**(6083): p. 897-901.

28. Saykally, R.J. and D.J. Wales. Chemistry. Pinning down the water hexamer. *Science*, 2012. **336**(6083): p. 814-5.

29. Moore, C. and S. Mertens, *The Nature of Computation*. 2011: Oxford University Press. p. 223-84.

30. Diaconis, P. The Markov chain Monte Carlo Revolution. *Bull Am Math Soc*, 2009. **46**(2): p. 179-205.

31. Chomsky, N. Language and Nature. *Mind*, 1995. **104**(413): p. 1-61.

32. Davies, D.D. and T.J. Humphrey. Amino Acid recycling in relation to protein turnover. *Plant Physiol*, 1978. **61**(1): p. 54-8.

33. Yewdell, J.W., et al. Out with the old, in with the new? Comparing methods for measuring protein degradation. *Cell Biol Int*, 2011. **35**(5): p. 457-62.

34. Weigel, A.V., et al. Ergodic and nonergodic processes coexist in the plasma membrane as observed by single-molecule tracking. *Proc Natl Acad Sci U S A*, 2011. **108**(16): p. 6438-43.

35. Pelkmans, L. Cell Biology. Using cell-to-cell variability--a new era in molecular biology. *Science*, 2012. **336**(6080): p. 425-6.

36. Leslie, M. Mysteries of the cell. Do lipid rafts exist? *Science*, 2011. **334**(6059): p. 1046-7.

37. Braakman, R. and E. Smith. The emergence and early evolution of biological carbon-fixation. *PLoS Comput Biol*, 2012. **8**(4): p. e1002455.

38. Pfaff, D. and J.R. Banavar. A theoretical framework for CNS arousal. *Bioessays*, 2007. **29**(8): p. 803-10.

39. Chung, H.S., et al. Single-molecule fluorescence experiments determine protein folding transition path times. *Science*, 2012. **335**(6071): p. 981-4.